\documentclass[aps,pra,twocolumn,showpacs]{revtex4}

\usepackage{graphicx}% Include figure files
\usepackage{dcolumn}% Align table columns on decimal point
\usepackage{bm}% bold math
\newcommand{\bra}[1]{\mathop{\left\langle #1 \right|}\nolimits}
\newcommand{\ket}[1]{\mathop{\left| #1 \right\rangle}\nolimits}
\newcommand{\Tr}{{\rm Tr \,}}

\begin{document}
\title{Mutual information--based approach to adaptive homodyne detection \\
of quantum optical states}
\author{Igor Bargatin}
\affiliation{Department of Physics 103-33, California Institute of Technology,
Pasadena, CA 91125}

\pacs{03.67.Hk, 42.50.Dv, 42.50.Lc}
\begin{abstract}
I propose an approach to adaptive homodyne detection of digitally modulated
quantum optical pulses in which the phase of the local oscillator is chosen to
maximize the average information gain, i.e., the mutual information, at each
step of the measurement. I study the properties of this adaptive detection
scheme by considering the problem of classical information content of
ensembles of coherent states. Using simulations of quantum trajectories and
visualizations of corresponding measurement operators, I show that the
proposed measurement scheme adapts itself to the features of each ensemble.
For all considered ensembles of coherent states, it consistently outperforms
heterodyne detection and Wiseman's adaptive scheme for phase measurements
[H.M. Wiseman, Phys. Rev. Lett. {\bf 75}, 4587 (1995)].
\end{abstract}
\maketitle

\section{Introduction}

Digital communication with modulated optical pulses is essential to the modern
interconnected world. While most optical communication schemes use strong
electromagnetic signals that are well described classically \cite{textbook},
some electromagnetic signals can exhibit manifestly non-classical behavior and
need to be analyzed using quantum mechanics \cite{wolf}. The latter situation
arises, for example, when electromagnetic signals are extremely weak either by
design, as in quantum cryptography \cite{crypt}, or out of necessity, as in
deep space communications \cite{space}.

Among the many types of quantum states of electromagnetic field that can be
used for communication are photon number states, coherent states, and
quadrature-squeezed states \cite{wolf}. Coherent states are especially popular
because they are quasiclassical in their properties and relatively easy to
prepare. There are also many different ways to measure electromagnetic states,
with direct photon counting, heterodyne detection, and homodyne detection
being easiest to implement in experiment. However, because of quantum
uncertainty, none of these methods can perfectly measure both quadratures of
the field. In fact, one can only decrease the measurement error in one
quadrature at the expense of an increase in the other \cite{wolf,cavesRMP}. In
heterodyne detection, the quickly rotating phase of the local oscillator
implies that all quadratures are measured equally well, which makes this
scheme the most versatile. Homodyne detection is the other extreme: It keeps
the local oscillator phase constant and therefore measures one quadrature
perfectly, while providing no information whatsoever about the perpendicular
one.

In this paper, I consider quantum detection of coherent states using homodyne
detection with the adaptively changing phase of the local oscillator
\cite{f1}. Allowing adaptive phase in homodyne measurements significantly
expands the set of possible quantum measurement of optical pulses, while its
experimental realization remains relatively straightforward. This type of
quantum measurement was proposed recently by Wiseman for various phase
measurement problems and was demonstrated to be superior to other types of
measurements both theoretically [see, for example, Ref. \cite{wiseman05} and
references therein] and experimentally \cite{hideoexp}.

There are many criteria for choosing a quantum measurement among different
alternatives. For example, one can try to minimize the probability of error in
determining the source variable from measurement results \cite{QAM,geremia},
the average squared deviation of the best estimate from the actual value of
the source variable \cite{wiseman95}, or maximize the mutual information
between the source variable and the measurement results \cite{QAM,cavesRMP}.
Optimization of these target functions is interrelated to some extent
\cite{osaki98}; for example, zero probability of error or zero deviation of
the estimate implies maximum mutual information and vice versa. However, in
this paper, I focus on maximizing the mutual information because it is the
mutual information that defines the information capacity of a communication
channel \cite{textbook,preskill}.

%The capacity of an information channel is the maximum rate at which
%information an be sent through the channel.

Note that, for quantum channels, one can define different types of classical
information capacities depending on whether one can perform collective quantum
measurements or only measure one state at a time, and whether communication is
allowed between measurement \cite{shor04}. While collective measurements
potentially result in a higher capacity, they highly impractical in the case
of continuous communication with coherent optical pulses. Performing adaptive
measurements of individual pulses is therefore an attractive way to extract
more information from a given state without resorting to advanced quantum
measurement techniques.

\section{Theory}

Let the state of the electromagnetic field be given by a coherent state
$\ket{\alpha_k}$ that depends on the value of the source variable $k$. I will
assume for simplicity that the source variable can only take a finite number
of values with known \emph{a priori} probabilities $p_k$. These probabilities
and states form an ensemble ${\cal E}=\left\{p_k,\ket{\alpha_k}\right\}$,
which appears in the problem of the classical information content of quantum
states \cite{preskill}. For efficient communication, one needs to maximize the
mutual information between the source variable and the measurement results by
optimizing the quantum measurement procedure. In general, this problem is not
solved, although optimal solutions are known for certain symmetric ensembles
of states \cite{symmetric}, and a ``pretty good'', but not necessarily
optimal, measurement can be derived for any ensemble \cite{pgm,QAM}. In this
paper, I consider only those measurements that can be realized using balanced
photodetection with an arbitrary time dependence of the local oscillator
phase. To the best of my knowledge, the optimal solution is not known in this
case either.

\begin{figure}
\includegraphics[width=3in]{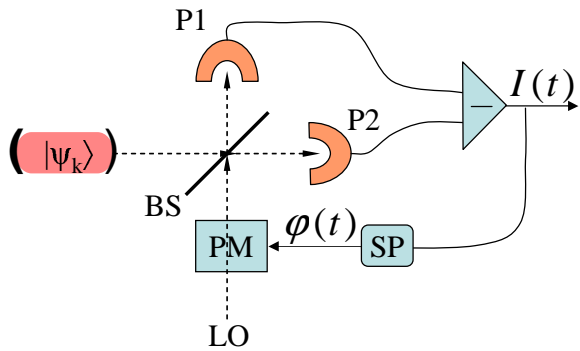}% Here is how to import EPS art
\caption{\label{fig:schematic} Schematic of adaptive homodyne measurement. The
signal processor (SP) analyzes the collected photocurrent record
$\{I(\tau):\,0<\tau<t\}$ and adjusts the phase of local oscillator (LO) using
the phase modulator (PM).}
\end{figure}

Figure \ref{fig:schematic} shows the relevant experimental setup, similar to
the one used recently to demonstrate improved optical phase estimation with
adaptive homodyne measurements \cite{hideoexp}. The optical cavity supports a
mode whose state is described by one of the wave functions from the ensemble
${\cal E}$ above. One of the mirrors of the otherwise lossless cavity is not
perfect, so the radiation leaks out and mixes with the strong beam of the
local oscillator (LO) at the 50-50 beam splitter (BS). In balanced
photodetection, one records the difference between the photocurrents of the
two detectors (P1,P2) in order to reduce the strong background due to the
local oscillator. The photocurrent record is then analyzed by the signal
processor (SP) to determine the optimal phase $\varphi$ of the local
oscillator (LO) for subsequent measurement. The phase can be updated
continuously or, more practically, discretely with a small time step $\Delta
t$. I will assume that there are no time delays in the feedback loop.

During a detection time interval $(t,t+\Delta t)$, each of photodetectors P1
and P2 generates a certain amount of electronic charge. In the following, I
will assume that the photodetectors are noiseless and perfectly efficient. If
the cavity state is initially given by a coherent state $\ket{\alpha_k}$ with
the complex amplitude $\alpha_k$, and the approximation of a strong local
oscillator applies, the difference between the generated charges can be
normalized to give \cite{wiseman95}
\begin{equation}\label{stoch}
  \Delta Q(t) = \int_t^{t+\Delta t}I(\tau)d\tau\approx 2 e^{-t/2}\Re[\alpha e^{-i\varphi(t)}]\Delta t + dW,
\end{equation}
where $\Re$ represents the real part of a number, $dW$ is the Wiener increment
of the quantum photodetection noise, satisfying $\left< dW^2\right>=\Delta t$,
and time has been scaled by the cavity decay time. The factor $e^{-t/2}$ in
Eq. (\ref{stoch}) describes amplitude decay of any coherent state in the
cavity due to leakage through the imperfect mirror. Each coherent state of the
field in the cavity therefore produces exponentially shaped pulses of
photocurrents starting at time $t=0$.

The probability distribution function for the accumulated photocharge is given
by \cite{wiseman97}
\begin{equation}\label{pdf}
  P\left(\Delta Q(t)\right) ={\cal N} \left( \Delta Q(t);2\Delta t\Re[\alpha e^{-t/2-i\varphi(t)}],\Delta
  t\right),
\end{equation}
where ${\cal N} \left(\Delta Q(t);A,\sigma^2
\right)\equiv\frac{1}{\sqrt{2\pi}\sigma}e^{-\frac{(\Delta
Q(t)-A)^2}{2\sigma^2}}$ is the Gaussian distribution with the mean $A$ and
standard deviation $\sigma$. For any measured value of the photocharge $\Delta
Q(t)$, one can update the prior probabilities of the source variable using
Bayes' rule:
\begin{eqnarray}\label{probBayes}
  p_k(t+\Delta t)=\frac{p_k(t){\cal N} \left(\Delta Q(t); 2\Delta t\Re[\alpha_k e^{-t/2-i\varphi(t)}],\Delta
  t\right)}{\sum_{k}p_k(t){\cal N} \left( \Delta Q(t);2\Delta t\Re[\alpha_k e^{-t/2-i\varphi(t)}],\Delta
  t\right)}.
\end{eqnarray}
For each result of the current measurement step, we therefore learn something
about the source variable. The gain in information (reduction in uncertainty)
about the source variable $k$ is given by
\begin{equation}\label{infgain}
  {\cal G}\,[\Delta Q(t)]=H(\left\{p_k(t)\right\})-H(\left\{p_k(t+\Delta
  t)\right\}),
\end{equation}
where $H(\left\{p_k\right\})=-\sum_k p_k\log_2 p_k$ is the Shannon entropy of
a probability distribution \cite{textbook,preskill}. The mutual information
between the measurement result $\Delta Q(t)$ and the source variable is simply
the information gain (\ref{infgain}) averaged over the random outcomes $\Delta
Q(t)$:
\begin{equation}\label{mutinf}
  {\cal I}\,[\Delta Q(t):k]=\left<H(\left\{p_k(t)\right\})-H(\left\{p_k(t+\Delta
  t)\right\})\right>_{\Delta Q(t)}.
\end{equation}

If the time period $\Delta t$ is sufficiently small, the standard deviation of
the photocharge probability distribution becomes much larger than its mean,
$\sigma\gg A$, because in Eq.~(\ref{pdf}), the mean scales as $\Delta t$,
whereas the standard deviation scales as $\sqrt{\Delta t}$ \cite{f2}. In this
case, the photocharge accumulated during one sampling period carries very
little information about the source variable, and only the totality of the
sampled photocharges may be sufficient to distinguish the states
$\ket{\alpha_k}$. Below I will consider only this situation because it arises
naturally in adaptive measurements.

The Gaussian distributions (\ref{pdf}), all having the same dispersion, are
then very wide and only slightly shifted from the zero mean. Expanding the
exponentials in $A$ up to the first order, we obtain
\begin{eqnarray}\label{explin}
{\cal N} \left(\Delta Q;A,\sigma^2
\right)%=\frac{1}{\sqrt{2\pi}\sigma}e^{-\frac{(\Delta Q-A)^2}{2\sigma^2}}
\approx \frac{1}{\sqrt{2\pi}\sigma}e^{-\frac{\Delta Q^2}{2\sigma^2}+\frac{\Delta Q}{\sigma^2}A}\\
\approx{\cal N} \left(\Delta Q;0,\sigma^2 \right)\left(1 +\frac{\Delta
Q}{\sigma^2}A \right).
\end{eqnarray}
The mutual information (\ref{mutinf}) is then approximately given by
\begin{equation}\label{inflin}
\begin{array}{l}
{\cal I}\,[\Delta Q(t):k]\approx\left<p_k{\cal N} \left(\Delta Q(t);0,\Delta
t\right)\left(1 +\frac{\Delta Q(t)}{\Delta t}A_k(t) \right)\times\right. \\
\left.\times\log \left(\frac{1 +\frac{\Delta Q(t)}{\Delta
t}A_k(t)}{\sum_{l}p_l\left[1 +\frac{\Delta Q(t)}{\Delta
t}A_l(t)\right]}\right)\right>_{k,\Delta Q}
%\approx\left<{\cal N} \left(\Delta Q(t);0,\Delta t\right)\frac{\Delta
%Q(t)^2\left[\left<A(t)^2\right>-\left<A(t)\right>^2\right]}{2\Delta
% t^2} \right>_Q=
\approx\frac{\left<A(t)^2\right>-\left<A(t)\right>^2}{2\Delta
 t},
\end{array}
\end{equation}
where in our case $A_k(t)=2 e^{-t/2}\Re[\alpha_k e^{-i\varphi(t)}]\Delta t$,
$\left<A(t)\right>=\sum_k p_k(t)A_k(t)$, and $\left<A(t)^2\right>=\sum_k
p_k(t)A_k^2(t)$.

Introducing new notation $X_k=\Re(\alpha_k)$ and $Y_k=\Im(\alpha_k)$, where
$\Im(.)$ is the imaginary part of a complex number, it is easy to show that
the last term in Eq. (\ref{inflin}) is proportional to
\begin{equation}\label{dispersion}
\begin{array}{l}
2\sigma^2_X(t)\cos^2\varphi+2\sigma^2_Y(t)\sin^2\varphi + 4
\sigma_{XY}(t)\sin \varphi\cos\varphi=\\
\sigma^2_X(t)+\sigma^2_Y(t)+ (\sigma^2_X(t)-\sigma^2_Y(t))\cos 2\varphi + 2
\sigma_{XY}(t)\sin 2\varphi,
\end{array}
\end{equation}
where $\sigma^2_{X}(t)\equiv\sum_k p_k(t)X_k^2-\left(\sum p_k(t)X_k\right)^2$,
$\sigma^2_{Y}(t)\equiv\sum_k p_k(t)Y_k^2-\left(\sum p_k(t)Y_k\right)^2$, and
$\sigma_{XY}\equiv\sum_k p_k(t)X_kY_k-\left(\sum p_k(t)X_k\right)\left(\sum
p_k(t)Y_k\right)$. Equation (\ref{dispersion}) is maximized when
\begin{equation}\label{sol}
 \varphi(t)=\frac{1}{2}{\rm
 Arg}\left[(\sigma^2_{X}(t)-\sigma^2_{Y}(t))+2i\sigma_{XY}(t)\right],
\end{equation}
where ${\rm Arg}[.]$ is the argument of a complex number.

Note that Eq. (\ref{inflin}) provides a simple geometrical interpretation of
the considered maximization problem. The ensemble $\{p_k, \ket{\alpha_k\}}$
defines an ensemble of points $\{p_k, X_k, Y_k\}$ in the phase space $XOY$.
These points can be projected onto a new coordinate axis $OP$, which forms an
angle $\varphi$ with axis $OX$, to produce a new ensemble $\{p_k, P_k\}$. The
dispersion of this ensemble $\{p_k, P_k\}$ is then proportional to the
expression (\ref{inflin}). In our maximization of mutual information, we are
therefore looking for a configuration that maximizes the expected dispersion
of the measured field quadrature at each measurement step.

The measurement scheme given by Eq. (\ref{sol}) is adaptive because the
probabilities $p_k(t)$ are updated according to Eqs. (\ref{probBayes}) after
each measurement step. The resulting detection scheme is locally optimal in
the sense that the average information gain is maximized at each measurement
step. For convenience, I will call this adaptive scheme LMMI measurement, from
Local Maximization of Mutual Information. Note that, even though local
optimization can sometimes lead to a globally optimal solution, it is not a
general rule and there is no guarantee that the LMMI measurement is optimal
globally, i.e., it maximizes the information gain from the entire measurement
record. However, I demonstrate below that the LMMI measurement is quite
versatile and, in all considered examples, performs better than heterodyne
detection and Wiseman's adaptive scheme for phase measurements.

Before presenting the results of numerical quantum trajectory simulations, it
is useful to review some limits on the information capacity of optical
communication channels \cite{cavesRMP}. For example, if one uses an ensemble
of coherent states and (nonadaptive) heterodyne detection, the mutual
information for a single pulse has an upper bound ${\cal
I}_1=\log_2\left(1+\left<n\right>\right)$ bits, where $\left<n\right>\equiv
\sum_k p_k|\alpha_k|^2$ is the average number of photons, i.e., the energy
used per pulse. This bound can be saturated using an infinite ensemble of
coherent states with the Gaussian distribution of prior probabilities. If one
allows squeezed states and homodyne detection, the upper bound increases to
${\cal I}_2=\log_2(1+2\left<n\right>)$. Presumably, this is the best one can
do with non-adaptive balanced photodetection, but the fragility of squeezed
states makes this bound difficult to reach in practice. Finally, another bound
on the mutual information is provided by the Holevo information of the given
ensemble of pure states $\chi({\cal E})=-\Tr{\hat\rho\log_2\hat\rho}$, where
$\hat\rho=\sum_k p_k\ket{\alpha_k}\bra{\alpha_k}$ \cite{preskill,cavesRMP}.
The Holevo information itself has a upper bound of ${\cal
I}_3=\log_2(1+\left<n\right>)+\left<n\right>\log_2(1+1/\left<n\right>)$. This
maximum capacity can be achieved using ensembles consisting of photon number
states with the Boltzmann distribution of prior probabilities and using
perfect photon counting for detection \cite{cavesRMP}. However, controlled
production of photon number states and perfect photon counting remain
technically challenging and are unlikely to become practical in the nearest
future.

\begin{figure}
\includegraphics[width=3.2in]{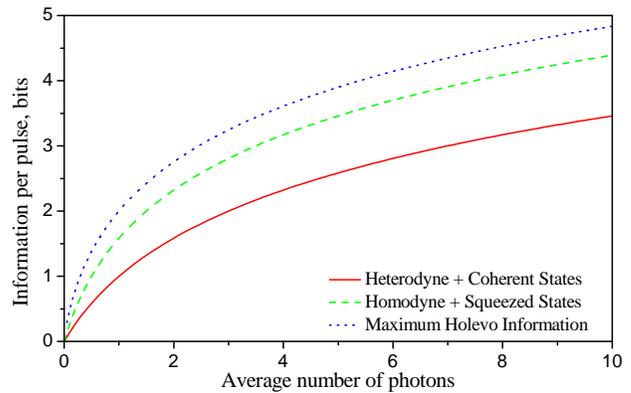}% Here is how to import EPS art
\caption{\label{fig:limits} Some general limits on the information capacity of
bosonic communication channels.}
\end{figure}

Figure \ref{fig:limits} shows all these bounds on a single graph. Note that
for strong signals, $\left<n\right>\gg 1$, these bounds are approximately
given by ${\cal I}_1\approx\log_2\left<n\right>$, ${\cal
I}_2\approx\log_2\left<n\right>+1$, and ${\cal
I}_2\approx\log_2\left<n\right>+\log_2 e\approx \log_2\left<n\right>+1.443$.
Therefore, in this semiclassical regime, the difference between the best
performance of heterodyne detection, ${\cal I}_1$, and the performance of the
best possible quantum measurement scheme, ${\cal I}_3$, is relatively small,
which probably explains the popularity of heterodyne detection in practical
applications. It is only in the limit of small photon numbers that one may
noticeably improve upon heterodyne detection with an adaptive measurement
scheme. Note also that the best energy efficiency of communication, i.e., the
amount of information transmitted per number of photons used, is achieved with
small-photon-number ensembles as well \cite{cavesRMP}.

\section{Numerical simulations}

\begin{figure}
\includegraphics[width=3.2in]{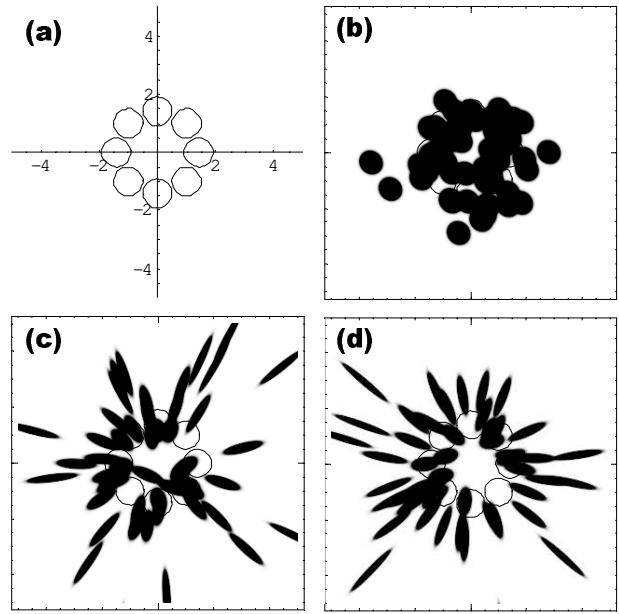}% Here is how to import EPS art
\caption{\label{fig:8psk} (a) Phase space representation of the ensemble of
eight equiprobable coherent states $\ket{\alpha_k}=\ket{\sqrt{2}e^{2\pi
ik/8}}, \, k=1..8$, with the same amplitude and evenly distributed phases
(8PSK). (b)-(d) The original ensemble and visualization of 50 projector
operators that represent POVM's corresponding to (b) (nonadaptive) heterodyne
measurement, (c) Wiseman's adaptive phase measurement, and (d) LMMI
measurement.}
\end{figure}

\begin{figure}
\includegraphics[width=3.2in]{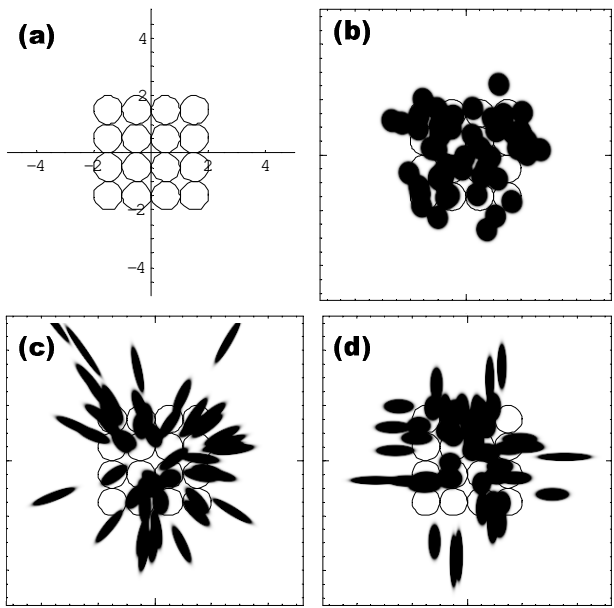}% Here is how to import EPS art
\caption{\label{fig:16qam} Same as Fig. \ref{fig:8psk} for an equiprobable
quadrature-amplitude modulated ensemble of 16 coherent states (16QAM).}
\end{figure}

To illustrate the performance of the proposed adaptive measurement scheme, I
have performed numerical quantum-trajectory simulations
\cite{plenio,wiseman_repres} for three different ensembles of relatively weak
coherent states with $\left<n\right><5$. Figures
\ref{fig:8psk}(a)--\ref{fig:star}(a) show these three ensembles on the $XOY$
phase plane, with the centers of circles $(X_k,Y_k)$ representing the
amplitudes of the corresponding states of the ensemble and the radius of each
circle representing the intrinsic quantum uncertainty of a coherent state
\cite{wolf,cavesRMP}.

The first ensemble (Fig. \ref{fig:8psk}(a)) consists of eight equiprobable
states with the same amplitude $|\alpha_k|=\sqrt{2}$ and evenly distributed
phases. In communications language \cite{textbook}, this modulation scheme is
known as phase-shift keying (PSK), and the ensemble is called 8PSK for short.
The second ensemble (Fig. \ref{fig:16qam}(a)) consists of 16 equiprobable
states with the real and imaginary parts of their amplitudes ranging from -1.5
to 1.5 with unit increment. This is so-called quadrature-amplitude modulation
(QAM) and the ensemble is called 16QAM. The third ensemble (Fig.
\ref{fig:star}(a)) consists of 10 equiprobable states arranged in the shape of
a three-lobe star, with all states having integer amplitudes from 0 to 3 and
phases of $0,\,2\pi/3,$ or $4\pi/3$. This type of combined phase and amplitude
modulation is not usually used in communications but illustrates well some
properties of the proposed adaptive measurement scheme. For convenience, I
will call it STAR.

The numerical simulations were performed using Mathematica software. The
discrete time step was $\Delta t=5\times10^{-3}$, so that $\alpha\Delta t\ll
\sqrt{\Delta t}$ for all simulates states. Each quantum trajectory was
simulated from $t=0$ to $t=10$, by which point the residual photon population
of the cavity is $e^{-10}<5\times 10^{-5}$ of the initial value, and the
information still extractable from it is negligible. For each ensemble, I
simulated a total of 10000 trajectories, randomly choosing an initial coherent
state from the given ensemble with the respective \emph{a priori}
probabilities and recording the total information gain ${\cal G}$ for each
trajectory. The statistic mean of the information gain and its standard
deviation were estimated from these data. They are presented for each ensemble
in Table \ref{results} in the form of the $2\sigma$-confidence intervals for
the mutual information between the source variable $k$ and the entire
photocharge record $\{\Delta Q(t):0<t<10\}$. The table also lists the average
number of photons per pulse for each ensemble $\left<n\right>$, the optimal
mutual information of heterodyne detection for this number of photons ${\cal
I}_1(\left<n\right>)$, and the Holevo information $\chi(\cal E)$ of each
ensemble. The latter was calculated by truncating the Hilbert space to the
maximum photon number of 100.

To provide benchmarks for the performance of the LMMI measurement, I have also
simulated quantum trajectories for a discrete approximation to heterodyne
detection, in which the phase of the local oscillator is increased by 0.1 rad
after each measurement step, and Wiseman's original adaptive phase measurement
scheme, in which the local oscillator phase is changed by $\Delta
Q(t)/\sqrt{t}$ after each measurement step \cite{wiseman95}. Note that
Wiseman's scheme was originally proposed to measure a continuously and
uniformly distributed random phase, and therefore may be ill-suited to the
problem of classical information extraction, especially for ensembles 16QAM
and STAR. However, Wiseman's scheme is the only adaptive homodyne detection
scheme widely discussed in the literature, and it is therefore instructive to
compare its properties to those of the adaptive LMMI measurement. Table
\ref{results} lists the estimated mutual information for all combinations of
the three considered detection techniques and the three ensembles. I will
defer the discussion of these results to the following sections.

\begin{table}
\begin{tabular}{l|c|c|c}
\hline\hline Ensemble ${\cal E}$&8PSK&16QAM&STAR\\
\hline  % put a line under headers
$\left<n\right>$&2&2.5&4.2\\
${\cal I}_1(\left<n\right>)$, bits&1.585&1.807&2.379\\
$\chi({\cal E})$, bits&2.449&2.859&2.751\\
${\cal I}_{het}({\cal E})$, bits&$1.492\pm0.008$&$1.743\pm0.011$&$1.872\pm0.009$\\
${\cal I}_{W}({\cal E})$, bits&$1.676\pm0.006$&$1.771\pm0.008$&$1.649\pm0.005$\\
${\cal I}_{LMMI}({\cal E})$, bits&$1.692\pm0.005$&$1.805\pm0.011$&$2.206\pm0.007$\\
\hline\hline
\end{tabular}
\caption{\label{results} Some properties of the considered ensembles and
results of numerical simulations for the average information gain (mutual
information) using heterodyne detection ${\cal I}_{het}({\cal E})$, Wiseman's
adaptive phase measurement ${\cal I}_W({\cal E})$, and adaptive LMMI
measurement ${\cal I}_{LMMI}({\cal E})$.}
\end{table}

\section{POVM visualizations}

Adaptive or not, any quantum measurement with a predetermined algorithm for
choosing the local oscillator phase can be represented as a generalized
quantum measurement known as the positive operator-valued measure (POVM)
\cite{preskill,wiseman_repres}. Wiseman has shown \cite{wiseman_repres} that,
for measurements with any time dependence of the local oscillator phase, these
POVMs consist of an infinite number of projectors onto pure squeezed states
$\ket{\alpha,\xi}$. To get additional insight in the properties of such
measurements, it is helpful to visualize a sample of these squeezed states on
the $XOY$ plane as ellipses that represent the density plot of the Wigner
functions of these states. Each ellipse corresponds to one quantum trajectory
(the outcome of one complete measurement) according to the equations
\begin{equation}\label{repres}
\alpha=\frac{A+BA^*}{1-|B|^2}, \quad \xi=\frac{-B\,{\rm arctanh} B}{|B|},
\end{equation}
where $A$ and $B$ are the following functionals of the entire photocurrent
measurement record \cite{wiseman97}:
\begin{eqnarray}\label{AB}
A=\int I(t) e^{i\varphi(t)-t/2}dt\approx \sum \Delta Q(t) e^{i\varphi(t)-t/2}\\
B=-\int e^{i2\varphi(t)-t}dt\approx -\sum 2e^{i\varphi(t)-t}\Delta t.
\end{eqnarray}

\begin{figure}
\includegraphics[width=3.2in]{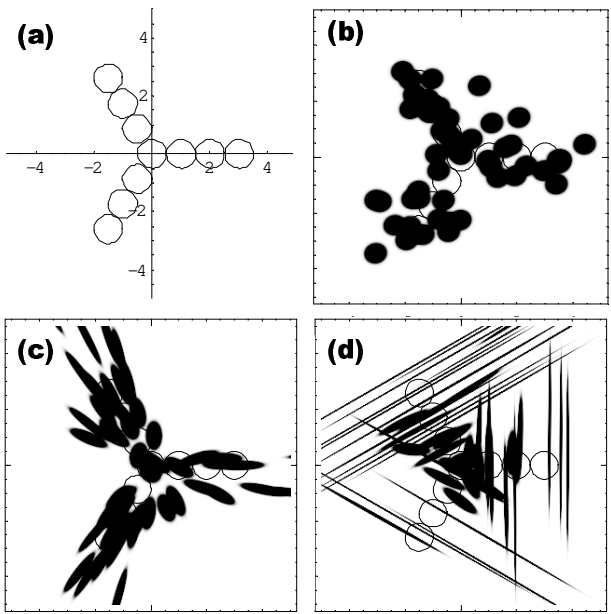}% Here is how to import EPS art
\caption{\label{fig:star} Same as Fig. \ref{fig:8psk} for an equiprobable
star-shaped ensemble of 10 coherent states (STAR).}
\end{figure}

Each POVM visualization in Figs. \ref{fig:8psk}--\ref{fig:star} shows fifty
such states, representing a random sample of the projectors that form the
corresponding POVM. Note that the projectors occur in this sample with the
same probabilities as the corresponding quantum trajectories and therefore
reflect the \emph{a priori} probabilities and quantum uncertainty of the
states that form each ensemble ${\cal E}$. While such visualizations are not
very rigorous, they do demonstrate which quadratures are given a preference
during each measurement and which uncertainties are minimized as a result.

For example, the visualizations of nonadaptive heterodyne measurements in
Figs. \ref{fig:8psk}(b)--\ref{fig:star}(b) contain only circles because
heterodyne detection samples all quadratures equally. The heterodyne POVM
therefore consists of projectors onto coherent states \cite{wiseman_repres},
which are a subset of squeezed states with the zero squeezing parameter,
$\xi=0$. The circles of these coherent states in POVM visualizations clutter
around the states of the original ensemble ${\cal E}$ because only those
projectors that have a significant overlap with the states of the original
ensemble are likely to appear in the visualized sample.

The visualizations of Wiseman's adaptive scheme (Figs.
\ref{fig:8psk}(c)--\ref{fig:star}(c)) also mostly consist of states that have
a significant overlap with the coherent states of the original ensemble, but
they are manifestly squeezed in the phase quadrature. This is expected, as the
scheme was designed to measure phase and therefore tries to reduce the phase
uncertainty of the measurement projectors. In the case of the 8PSK ensemble,
this is quite appropriate, and Wiseman's scheme produces a significantly
larger average information gain than heterodyne detection, even surpassing the
limit of optimal heterodyne detection ${\cal I}_1(2)\approx 1.585$ [see Table
\ref{results}]. Interestingly enough, the POVM visualization of the LMMI
measurement (Fig. \ref{fig:8psk}(d)) also consist of phase-squeezed states and
looks very similar to that of Wiseman's scheme. It is therefore not surprising
that the average information gains of Wiseman's and LMMI schemes are almost
equal. The small advantage of the LMMI scheme probably stems from the fact
that the phases are not continuously distributed over $2\pi$, as in the
original derivation of Wiseman's scheme \cite{wiseman95}, but rather assume a
number of discrete values.

In the case of the 16QAM ensemble, the visualization of Wiseman's and LMMI
schemes look quite different (Figs. \ref{fig:16qam}(c) and
\ref{fig:16qam}(d)). Wiseman's scheme is at a disadvantage here because it, as
always, tries to do the best phase measurements by using phase-squeezed
states. Nevertheless, it still performs better than heterodyne detection
because this ensemble has a lot of information encoded in the phase of the
constituent coherent states. The visualization of the LMMI scheme is more
interesting. It consists of states squeezed predominantly in either $X$ or $Y$
direction, which obviously reflects the symmetry of the 16QAM ensemble. In the
course of a single measurement, the LMMI scheme first tries to determine the
general area in which the measured state is located and then performs either X
or Y homodyne measurement, whichever is more appropriate, in the remaining
time. Clearly, this approach pays off as the LMMI scheme results in a
statistically significant lead in mutual information over both heterodyne and
Wiseman's detection schemes.

Finally, in the case of the STAR ensemble, the differences between the three
visualizations are even more striking. As usual, the heterodyne scheme's
coherent states are scattered around the three lobes of the original ensemble
(Fig. \ref{fig:star}(b)). So are the phase-squeezed states of Wiseman's scheme
(Fig. \ref{fig:star}(c)), but they are so elongated in the radial direction
that they are almost incapable of distinguishing different states from the
same lobe. Interestingly, the average information gain of Wiseman's scheme is
only slightly larger than $\log_2 3\approx 1.585$, which may be interpreted to
result from perfect discrimination of the phase of each coherent state, but
very poor discrimination in the amplitude of coherent states from the same
lobe.

The visualization of the LMMI measurement of the STAR ensemble predominantly
consists of highly squeezed states that are perpendicular to the three lobes
of the ensemble. The LMMI scheme therefore first quickly determines the phase
of a given state, and then performs homodyne measurement of its amplitude. As
a result, it performs noticeably better than both the heterodyne scheme and
Wiseman's scheme, covering a significant fraction of information gap between
the performances of the heterodyne scheme and the best quantum measurement, as
specified by the Holevo information of the ensemble.

\section{Discussion and conclusions}

The purpose of this study was to present a new adaptive detection technique
and explore some of its properties rather than conduct an exhaustive numerical
analysis of its performance. Therefore, I presented the simulation results for
only three ensembles that were specifically chosen to highlight the properties
of the considered measurement schemes. In all of them, the LMMI scheme
statistically outperforms both the heterodyne and Wiseman's schemes.
Nevertheless, this seems to be a general result. While I have performed
similar simulations with other ensembles of coherent states, I have never
found an ensemble where the LMMI scheme would perform worse than either the
heterodyne or Wiseman's scheme.

It is clear from the discussion above that adaptive measurements generally
make a good use of prior measurement results in determining the optimal local
oscillator phase. Wiseman's scheme is designed to measure the phase and
therefore performs particularly well with phase-modulated ensembles. The LMMI
scheme is more versatile in that it can measure the phase as well as Wiseman's
scheme but can also perform other types of adaptive homodyne measurements when
the ensemble features call for it.

As a price for its better performance, the LMMI scheme is much more demanding
computationally, as probabilities have to be updated after each step and the
new phase calculated according to the relatively complicated Eq. (\ref{sol}).
This reflects a traditional tradeoff between information capacity and
computational complexity that is typical of many communication problems.
However, with the ever increasing speed and decreasing cost of computing
power, the general trend has recently been towards more sophisticated schemes
that can extract more information from imperfect channels.

The analysis presented in this paper can be relatively easily generalized to
include the case of unequal \emph{a priori} probabilities and noisy
photodetectors, but the results are qualitatively similar to the ones
discussed above. Note that adaptive schemes generally seem to be more robust
with respect to instrumental imperfections than nonadaptive ones
\cite{geremia}. In the future, it would be interesting to extend the analysis
to squeezed states and determine whether it is possible to beat the homodyne
limit ${\cal I}_2(\left<n\right>)$ with adaptive measurements of ensembles of,
for example, phase-squeezed states. It would also be instructive to prove or
disproof the global optimality of the LMMI measurement scheme, but like any
nonlinear global optimization problem, it is probably a difficult task.

In conclusion, I have studied extraction of classical information from an
ensemble of coherent states using a new adaptive measurement scheme that
maximizes the average information gain (mutual information) at each step of
the adaptive measurement. Judging from the three considered examples, the
proposed LMMI scheme is quite versatile and adapts the measurement process to
the features of each ensemble. As a result, the LMMI scheme consistently
outperforms heterodyne detection and Wiseman's adaptive scheme. In the case of
the 16QAM and STAR ensembles, the improvement in extracted information with
respect to the heterodyne scheme was 4\% and 18\%, respectively. In the case
of the 8PSK ensemble, the average information gains of Wiseman's and LMMI
adaptive schemes are almost equal and about 13\% larger than the average
information gain of the heterodyne detection. In the latter case, the two
adaptive schemes even surpass the limit of heterodyne detection for the same
average number of photons per pulse. Compared to Wiseman's adaptive scheme,
the LMMI scheme is more computationally intensive, but its superior
performance and versatility may well justify its use for ensembles of weak
coherent states.

\section*{Acknowledgements}

I thank Hideo Mabuchi for comments on the manuscript and wish to acknowledge
the financial support from Caltech's Institute for Quantum Information during
the summer of 2001, when the bulk of this work was done.


\begin{thebibliography}{99}
\bibitem{f1}
In principle, there is no difference between continuous adaptive homodyne and
continuous adaptive heterodyne measurements since both allow an arbitrary
dependence of local oscillator phase on the prior detection record. I call the
measurements considered in this paper homodyne only because in the discrete
realizations simulated below, the detection is homodyne during each step of
the measurement.
\bibitem{f2}
Of course, $\Delta t$ should not be so small that the discrete nature of
photocounts begins to play a role. This can be avoided if the local oscillator
is sufficiently strong.
\end{thebibliography}
\end{document}